# The Influence of Approximation in Generalized Uncertainty Principle on Black Hole Evaporation

Xin-Dong Du and Chao-Yun Long[*]

Department of Physics, Guizhou University, Guiyang 550025, China

**Abstract**

The generalized uncertainty principle is often used to modify various thermodynamics systems by regarding the greater-than-equal relation as an approximate relation. We give a method to improve this approximation and compare the differences between the original and improved methods during the evaporation of black hole from two aspects of positive and negative parameters. Finally, we prove the rationality of the improved method and give some guiding opinions.

## 1. Introduction

Hawking radiation [1] supports a semi-classical evaporation of black hole: as the thermal emission goes on, the mass begins to decrease, so that the black hole gets smaller and hotter and radiates faster. In the end, the temperature drives to infinity and the black hole breaks down to its eventual disappearance. However, these results will be defective when quantum gravitational effects become leading [2]. The generalized uncertainty principle is a corrected result for the Heisenberg uncertainty principle by introducing the existence of a minimal length [3], and it can be used to modify the Hawking evaporation. The generalized uncertainty principle can prevent black hole from ending up with an explosion and remove the divergence of the final temperature of black hole [4].

The entropy can reflect the evaporation process, and the most common method to modify the entropy by the generalized uncertainty principle is from [5]. The common method is to directly change the sign of greater-than-equal to a sign of approximate equal and it is widely applied in [3, 6-11]. However, this approximation used in the method will cause huge impact on the evaporation of black hole, so that some thermodynamic variations in the evaporation are inconsistent with existing papers [12-18]. In this paper, based on [19], we give an improved approximation method to avoid the impact brought by the original method. In order to explain why the improved method is better than the original method, we will show the specific calculation processes and apply the two methods into the evaporation of black hole for comparison. In addition, quantum gravity [12] and non-commutative geometry [13], as other theories reflecting the minimum length, also prove the rationality of the improved method on the other hand.

[*]Corresponding author. E-mail: LongCYGuiZhou@163.com

If there is a remarkable influence caused by the approximation in a basic thermodynamic system, it will be hard to believe the approximation has no effect on a more complicated system. For this reason, the simplest Schwarzschild black hole is considered in this paper, and these involved thermodynamic quantities are all from Hawking radiation. As a standard to determine whether the evaporation of black hole is reasonable, the results of [14] will be used: a black hole's evaporation, tunneling to a white hole (remnant) [20], and final slow decay, form a unitary process that allows quantum gravity could resolve the information problem. In order to prevent the simplifications of calculations from interfering with the final results, Taylor expansion will not be applied to simplify any formula, and all physical constants will be reduced to 1 only when figures are painted.

## 2. The Difference between Two Approximation Methods

We first give the same initial steps for the two approximation methods. The metric linear element of Schwarzschild black hole space-time is given by:

$$ds^2 = -\left(1 - \frac{2GM}{c^2 r}\right) c^2 dt^2 + \left(1 - \frac{2GM}{c^2 r}\right)^{-1} dr^2 + r^2 d\Omega_2^2, \tag{1}$$

so the location of the Schwarzschild black hole horizon $r_H$ is:

$$r_H = \frac{2GM}{c^2}, \tag{2}$$

where $M$ is the mass of the black hole, $c$ is the speed of light, $r$ is the distance from the center of black hole, $t$ is the time, and $d\Omega_2^2$ is the standard metric on a two-dimensional spherical surface.

The modifications of the Heisenberg uncertainty principle due to the existence of a minimal length have been discussed by various theoretical models of gravity and quantum mechanics [21-23]. And string theoretical researches in Gedanken experiments [24-26] suggest that the Heisenberg uncertainty principle should be turned into the generalized uncertainty principle:

$$\Delta x \Delta p \geq \frac{\hbar}{2}\left(1 + \frac{\alpha L_p^2}{\hbar^2}\Delta p^2\right), \tag{3}$$

where $\Delta x$ and $\Delta p$ are the uncertainties of position and momentum, $\alpha$ is a dimensionless constant, and $L_p = \sqrt{G\hbar/c^3}$ is the Planck length.

Transform the right-hand part of Eq. (3) into a perfect square trinomial of $\Delta p$ and take its square root to obtain:

$$\frac{\Delta x \hbar}{\alpha L_p^2}\left[1 + \sqrt{1 - \frac{\alpha L_p^2}{(\Delta x)^2}}\right] \geq \Delta p \geq \frac{\Delta x \hbar}{\alpha L_p^2}\left[1 - \sqrt{1 - \frac{\alpha L_p^2}{(\Delta x)^2}}\right]. \tag{4}$$

Because the right-hand part of Eq. (4) can be reduced into the Heisenberg uncertainty principle as $\alpha \to 0$ while the left-hand part cannot do that, the right-hand part should be chosen instead of Eq. (4):

$$\Delta p \geq \frac{\Delta x \hbar}{\alpha L_p^2}\left[1 - \sqrt{1 - \frac{\alpha L_p^2}{(\Delta x)^2}}\right]. \tag{5}$$

Two approximation methods will be obtained according to the different treatments of Eq. (5). The first one is the most common method [5]: to directly change the sign of greater-than-equal of Eq. (5) into a sign of approximate equal. By contrast, the second one is what we came up with based on [19] and has a more accurate improvement: to regard Eq. (5) as the lower bound of $\Delta p$. For this reason, we call the first one the original method and the second one the improved method. In order to show the influences of these two methods on the evaporation of black hole, to start with, we will employ them separately to calculate the differential of black hole entropy:

## 2.1 The original method

The subscript A is used to represent the physical quantities modified by the original method. Then from Eq. (5) the corrected uncertainty of the momentum is:

$$\Delta p_A \simeq \frac{\Delta x \hbar}{\alpha L_p^2}\left[1 - \sqrt{1 - \frac{\alpha L_p^2}{(\Delta x)^2}}\right]. \tag{6}$$

According to [5], combining with Eq. (2), we should choose:

$$\Delta x \simeq r_H = \frac{2GM}{c^2}, \tag{7}$$

therefore,

$$\Delta p_A \simeq \frac{2Mc}{\alpha}\left[1 - \sqrt{1 - \frac{\alpha M_p^2}{4M^2}}\right], \tag{8}$$

where $M_p = \sqrt{\hbar c / G} = c^2 L_p / G$ is the Planck mass.

In [5], by introducing a calibration factor $1/2\pi$, the Hawking temperature of black hole is written as:

$$T = \frac{1}{2\pi}\frac{\Delta p c}{k_B}. \tag{9}$$

Because the differential of black hole entropy $dS = (c^2/T)dM$, the relationship between $dS$ and $\Delta p$ can be gained:

$$dS = \frac{2\pi k_B c}{\Delta p}dM. \tag{10}$$

Substituting Eq. (8) into Eq. (10), we can get:

$$dS_A \simeq \frac{4\pi k_B M}{M_p^2}\left[1 + \sqrt{1 - \frac{\alpha M_p^2}{4M^2}}\right]dM, \tag{11}$$

where $dS_A$ is the differential of black hole entropy obtained from the original method?

### 2.2 The improved method

The subscript G is used to represent the physical quantities modified by the improved method. Then from Eq. (5) the corrected minimum uncertainty of the momentum is:

$$(\Delta p_G)_{min} = \frac{\Delta x \hbar}{\alpha L_p^2}\left[1 - \sqrt{1 - \frac{\alpha L_p^2}{(\Delta x)^2}}\right]. \tag{12}$$

The main idea of improved method is to replace Eq. (6) with Eq. (12). The following calculations are for the Schwarzschild black hole not for other thermodynamic systems, and different systems require different treatments to apply Eq. (12), which is not the focus of this paper.

However, Eq. (6) and Eq. (12) are so similar in form and the only difference between them is whether a minimum limit for the corrected uncertainty of the momentum is contained, so we need to further clarify their connection and difference.

First Eq. (6) gives $\Delta p_A \simeq \Delta x \hbar / (\alpha L_p^2)\left[1 - \sqrt{1 - \alpha L_p^2/(\Delta x)^2}\right]$ that is always a decreasing function of $\Delta x$ whether $\alpha$ is positive or negative. For $\alpha < 0$,

$(\Delta p_A)_{min} \simeq 0$ as $\Delta x \to \infty$. For $\alpha > 0$, $(\Delta p_A)_{min} \simeq \hbar / (\sqrt{\alpha} L_p)$ as $(\Delta x)_{max} = \sqrt{\alpha} L_p$. In either case, the result from Eq. (6) completely disagrees with Eq. (12) (Eq. (12) $\not\Leftrightarrow$ Eq. (6)), because Eq. (6) shows the lower limit of $\Delta p_A$ is only a constant independent of $\Delta x$ while Eq. (12) shows the lower limit should be a function closely related to $\Delta x$. Thus, there is an essential difference between Eq. (6) and Eq. (12) even though they are so similar in form.

In addition, both Eq. (6) and Eq. (12) are from Eq. (5) and they are the cores of the original method and the improved method respectively, so understanding the relationships between them and Eq. (5) can help us to distinguish the improved method from the original method better. On the one hand, it can be clearly seen that Eq. (12) and Eq. (5) are formally equivalent, so there are no approximations arising when Eq. (5) is changed into Eq. (12). On the other hand, because Eq. (12) $\Leftrightarrow$ Eq. (5) and Eq. (12) $\not\Leftrightarrow$ Eq. (6), so Eq. (5) $\not\Leftrightarrow$ Eq. (6). It means there is an obvious approximation arising when Eq. (5) is changed into Eq. (6). From the above two aspects, we can see Eq. (12) reflects Eq. (5) well but Eq. (6) fails in that, which suggests Eq. (12) successfully inherits the characteristics of the generalized uncertainty principle while Eq. (6) loses most features. In other words, the obvious approximation in Eq. (6) will make the results of the original method deviate from the reasonable rang corrected by the generalized uncertainty principle, but the improved method can avoid the deviation well by using Eq. (12).

According to [27], combining with Eq. (2), we should choose:

$$\Delta x \simeq 2r_H = \frac{4GM}{c^2}, \tag{13}$$

therefore,

$$(\Delta p_G)_{min} \simeq \frac{4Mc}{\alpha}\left[1 - \sqrt{1 - \frac{\alpha M_p^2}{16M^2}}\right]. \tag{14}$$

The Heisenberg uncertainty principle $\Delta x \Delta p \geq \hbar / 2$ gives:

$$(\Delta p)_{min} = \frac{\hbar}{2\Delta x} \simeq \frac{\hbar c^2}{8GM} = \frac{M_p^2 c}{8M}, \tag{15}$$

where $(\Delta p)_{min}$ is the uncorrected minimum uncertainty of the momentum.

The area of the black hole horizon $A$ can be expressed as:

$$A = 4\pi r_H^2 = \frac{16\pi G^2 M^2}{c^4}, \tag{16}$$

so

$$dA = \frac{32\pi G^2 M}{c^4} dM, \qquad (17)$$

$$\Delta A = \frac{32\pi G^2 M}{c^4} \Delta M. \qquad (18)$$

For black hole absorbing or releasing particle of energy, we have:

$$\Delta M \simeq c\Delta p. \qquad (19)$$

Substituting Eq. (19) into Eq. (18), we can get:

$$\Delta A \simeq \frac{32\pi G^2 M}{c^3} \Delta p. \qquad (20)$$

According to the definition of differential, the differential of function $Y(X)$ can be expressed as $dY = \lim_{\Delta X \to 0} |Y(X + \Delta X) - Y(X)| = \lim_{\Delta X \to 0} \Delta Y \simeq (\Delta Y)_{min}$. From Eq. (16) the area can be regarded as the function on mass, so for $A_G(M)$ and $A(M)$(the revised area and the uncorrected area), we have $dA_G \simeq (\Delta A_G)_{min}$ and $dA \simeq (\Delta A)_{min}$. Divide the former by the latter to get: $dA_G / dA \simeq (\Delta A_G)_{min} / (\Delta A)_{min}$. Thus, combining with Eq. (20), we can obtain an approximate expression for the revised differential of the area $dA_G$:

$$dA_G \simeq \frac{(\Delta A_G)_{min}}{(\Delta A)_{min}} dA \simeq \frac{\frac{32\pi G^2 M}{c^3}(\Delta p_G)_{min}}{\frac{32\pi G^2 M}{c^3}(\Delta p)_{min}} dA = \frac{(\Delta p_G)_{min}}{(\Delta p)_{min}} dA. \qquad (21)$$

It is worth noting that Eq. (21) plays a so essential role to support our paper that we must show its influence clearly. On the one hand, it is through Eq. (21) that the core of the improved method Eq. (12) can be applied to calculate the following various thermodynamic quantities, so Eq. (21) supports the improved method in another way. On the other hand, if there is no limit to the minimum of the uncertainties of momentum, the core of the original method Eq. (6) can be also applied into Eq. (21) to lead to a same result as the improved method. Hence, the existence of Eq. (21) is very necessary to distinguish the improved method from the original method.

Connecting with Eqs. (14), (15) and (17), we can turn Eq. (21) into:

$$dA_G \simeq \frac{32M^2}{\alpha M_p^2}\left[1 - \sqrt{1 - \frac{\alpha M_p^2}{16M^2}}\right]\frac{32\pi G^2 M}{c^4} dM. \qquad (22)$$

Based on Bekenstein-Hawking area law $S = k_B A / (4L_p^2)$, Eq. (22) is rewritten as:

$$dS_G \simeq \frac{256\pi k_B M^3}{\alpha M_p^4}\left[1 - \sqrt{1 - \frac{\alpha M_p^2}{16M^2}}\right]dM, \qquad (23)$$

where $dS_G$ is the differential of black hole entropy obtained from the improved method.

## 3. Thermodynamic Quantities during The Evaporation

In order to show the changes of thermodynamic quantities during the evaporation of black hole, we shall give the functions of each thermodynamic quantity on the mass of black hole. And for every thermodynamic quantity, they all will have two different functions coming from the original and improved methods:

### 3.1 Entropy

Integrate both sides of Eq. (11) and Eq. (23) to yield:

$$S_A \simeq \frac{\alpha \pi k_B}{4}\left[\frac{2}{1 - \sqrt{1 - \frac{\alpha M_p^2}{4M^2}}} + \ln\left(1 - \sqrt{1 - \frac{\alpha M_p^2}{4M^2}}\right) - \ln\left(1 + \sqrt{1 - \frac{\alpha M_p^2}{4M^2}}\right)\right], (24)$$

$$S_G \simeq \frac{\alpha \pi k_B}{16}\left[\frac{2}{1 - \sqrt{1 - \frac{\alpha M_p^2}{16M^2}}} - \ln\left(1 - \sqrt{1 - \frac{\alpha M_p^2}{16M^2}}\right) + \ln\left(1 + \sqrt{1 - \frac{\alpha M_p^2}{16M^2}}\right)\right.$$

$$\left. + \frac{2}{\left(1 + \sqrt{1 - \frac{\alpha M_p^2}{16M^2}}\right)^2}\right]. \qquad (25)$$

It must be noted that the above results are obtained with a positive parameter and if the parameter is negative, the following change should be made for $K = 4, 16$:

$$\ln\left(1 - \sqrt{1 - \frac{\alpha M_p^2}{KM^2}}\right)(\alpha > 0) \Rightarrow \ln\left(\sqrt{1 - \frac{\alpha M_p^2}{KM^2}} - 1\right)(\alpha < 0). \qquad (26)$$

## 3.2 Temperature

Import Eq. (11) and Eq. (23) into $T = c^2 dM / dS$ to gain:

$$T_A = c^2 \frac{dM}{dS_A} \simeq \frac{Mc^2}{\pi \alpha k_B} \left[ 1 - \sqrt{1 - \frac{\alpha M_p^2}{4M^2}} \right], \tag{27}$$

$$T_G = c^2 \frac{dM}{dS_G} \simeq \frac{M_p^2 c^2}{16\pi k_B M} \left[ 1 + \sqrt{1 - \frac{\alpha M_p^2}{16M^2}} \right]. \tag{28}$$

## 3.3 Heat capacity

Take the derivatives of Eq. (27) and Eq. (28) with respect to $M$:

$$\frac{dT_A}{dM} \simeq \frac{c^2}{\pi \alpha k_B} \left[ 1 - \frac{1}{\sqrt{1 - \frac{\alpha}{4}\left(\frac{M}{M_p}\right)^{-2}}} \right], \tag{29}$$

$$\frac{dT_G}{dM} \simeq \frac{c^2}{16\pi k_B} \left[ -\left(\frac{M}{M_p}\right)^{-2} + \frac{-\left(\frac{M}{M_p}\right)^{-2} + \frac{\alpha}{8}\left(\frac{M}{M_p}\right)^{-4}}{\sqrt{1 - \frac{\alpha}{16}\left(\frac{M}{M_p}\right)^{-2}}} \right]. \tag{30}$$

Substituting the above two equations into $C = c^2 dM / dT$, we can get:

$$C_A = c^2 \frac{1}{\frac{dT_A}{dM}} \simeq -4\pi k_B \left[ \left(\frac{M}{M_p}\right)^2 - \frac{\alpha}{4} + \left(\frac{M}{M_p}\right)^2 \sqrt{1 - \frac{\alpha}{4}\left(\frac{M}{M_p}\right)^{-2}} \right], \tag{31}$$

$$C_G = c^2 \frac{1}{\frac{dT_G}{dM}} \simeq 16\pi k_B \frac{\left(\frac{M}{M_p}\right)^2}{\left[ -1 + \frac{-\left(\frac{M}{M_p}\right)^2 + \frac{\alpha}{8}}{\left(\frac{M}{M_p}\right)^2 \sqrt{1 - \frac{\alpha}{16}\left(\frac{M}{M_p}\right)^{-2}}} \right]}. \tag{32}$$

## 3.4 Evaporation rate

Via the Stefan-Boltzmann law [28], the radiated power is derived:

$$\frac{dE}{dt} = \sigma A T^4, \tag{33}$$

where $\sigma = \pi^2 k_B^4 / (60 \hbar^3 c^2)$ is known as the Stefan-Boltzmann constant [29]. But usually Eq. (33) should be modified by the squeezing of the phase space due to the generalized uncertainty principle, and the corrected result is given by [30]:

$$-\frac{dE}{dt} = 12\pi \Gamma_\gamma \sigma_3 R_S{}^2 A(\alpha,T) T^4, \tag{34}$$

where the minus sign represents the loss of mass or energy, $\Gamma_\gamma$ is the greybody factor, $\sigma_3$ is the Stefan-Boltzmann constant (in four dimensions), $R_S$ is the radius of the event horizon, and $A(\alpha,T)$ is a function about the parameter of generalized uncertainty principle and the temperature of black hole accounting for the cells' squeezing in momentum space.

The main difference between Eq. (33) and Eq. (34) is that $A$ in Eq. (33) is independent of $T$ while $A(\alpha,T)$ in Eq. (34) is a function related to $T$, which will exert an effect on the evaporation rate that we are going to calculate with $T$. However, the effect can be neglected for the following two reasons: first by using Planck variables, we can transform $A(\alpha,T)$ into $\mathcal{A}(\alpha,\Theta)$ ($\Theta = T/T_p$) and the value range of $\mathcal{A}(\alpha,\Theta)$ is derived by [30] as $0 < \mathcal{A}(\alpha,\Theta) \leq 1$. That means the value of $A(\alpha,T)$ does not fluctuate significantly as $T$ changes so that the influence caused by $A(\alpha,T)$ should be limited in a fixed range. For another, in [30], the figures of black hole mass with respect to time are shown to compare the cases including and not including the squeezing in the generalized uncertainty principle. From these figures, we can see mass thresholds are not at all affected by this squeezing. These figures also show the evaporation rates including the squeezing are smaller than that not including the squeezing, especially in the last phases of the evaporation, but the overall variation trends for the evaporation rates are consistent (In both cases, the evaporation rates increase with time). The analyses of the two aspects indicate the overall characteristics of the evaporation rate will not be drastically corrected by the squeezing. Thus, to some extent the phase space squeezing can be neglected in the calculation of evaporation rate, and we will use Eq. (33) instead of Eq. (34) for a much simpler and clearer procedure.

We assume the lost mass of black hole is completely converted to energy:

$$dE = c^2 dM. \tag{35}$$

From Eqs. (16), (33) and (35), the evaporation rate of black hole is obtained:

$$\frac{dM}{dt} = \frac{4}{15}\frac{\pi^3 k_B{}^4 G^2 M^2}{\hbar^3 c^8}T^4. \tag{36}$$

Substituting Eq. (27) and Eq. (28) into Eq. (36), we can gain:

$$\left(\frac{dM}{dt}\right)_A \simeq \frac{4}{15}\frac{\pi^3 k_B{}^4 G^2 M^2}{\hbar^3 c^8}T_A{}^4 \simeq \frac{4}{15}\frac{M^6 c^2}{\pi\alpha^4 \hbar M_p{}^4}\left[1-\sqrt{1-\frac{\alpha M_p{}^2}{4M^2}}\right]^4, \tag{37}$$

$$\left(\frac{dM}{dt}\right)_G \simeq \frac{4}{15}\frac{\pi^3 k_B{}^4 G^2 M^2}{\hbar^3 c^8}T_G{}^4 \simeq \frac{1}{60\times 16^3}\frac{M_p{}^4 c^2}{\pi\hbar M^2}\left[1+\sqrt{1-\frac{\alpha M_p{}^2}{16M^2}}\right]^4. \tag{38}$$

## 4. Evaporation under Parameters with Different Signs

The original and improved methods have given two different functions for every thermodynamic quantity in Section 3. In this section, we are going to draw the two functions of the same thermodynamic quantity on a common graph to compare their differences. The letter A and G will be used to respectively stand for the results caused by the original and improved methods. From here onwards, $k_B = c = \hbar = G = \pi = 1$ will be set to simplify the drawing of functions. In order to unify the remnant masses (critical masses) coming from the two methods, we take $|\alpha| = 2$ for the original method and $|\alpha| = 8$ for the improved method.

However, the sign of parameter of generalized uncertainty principle will lead to different results for the evaporation of black hole [8]. Normally the parameter is positive [31], but various scenarios have already proved negative parameter makes some sense such as an uncertainty relation from a crystal-like universe [32], the Magueijo and Smolin formulation of doubly special relativity [33] and a method to restore the Chandrasekhar limit corrected by the generalized uncertainty principle [34]. Thus, for the integrality of comparison, we are going to discuss the changes of thermodynamic quantities from positive and negative aspects:

**4.1 The evaporation with a positive parameter**

In figure 1, the decreasing trends of $S_A$ and $S_G$ are shown when $M$ decays. When $M$ arrives to the remnant mass, both $S_A$ and $S_G$ will become finite nonzero values, which agrees with [10]. However, the entropy should be 0 at the eventual evaporation stage [35], so the appearance of finite nonzero values suggests the evaporation can keep happening at Planck scale.

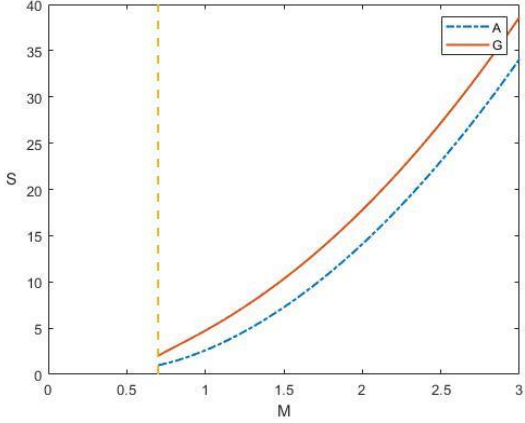

**Figure 1**. The entropy of black hole as given by Eq. (24) and Eq. (25) as the functions on $M$ for the original (A) and improved (G) methods when $\alpha > 0$.

In figure 2, $T_A$ continues to rise while $T_G$ goes up to a peak and down. Both of them finally arrive to finite nonzero temperatures, which confirms that the generalized uncertainty principle can remove the divergence of temperature [36]. Some papers [14, 15] argue that the temperature of remnant would be smaller than that of Hawking radiation, so that the remnant might slowly decay and quantum gravity could resolve the information problem. Thus, we consider $T_G$ is more reasonable than $T_A$ for the evaporation of remnant.

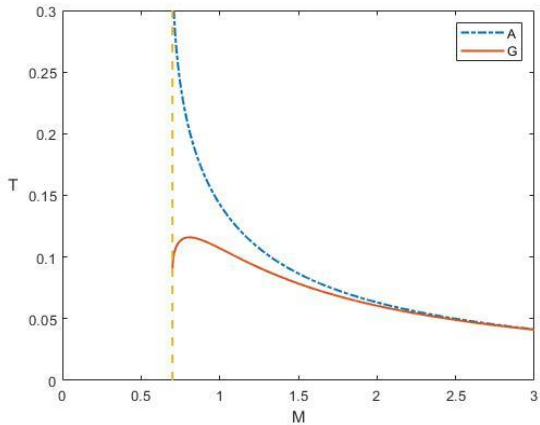

**Figure 2**. The temperature of black hole as given by Eq. (27) and Eq. (28) as the functions on $M$ for the original (A) and improved (G) methods when $\alpha > 0$.

In figure 3, $C_A$ rises from negative value to 0. By contrast, $C_G$ also become 0 in the last, but it is positive near the remnant mass and negative in the other, and between them $C_G$ diverges. The variation trend of $C_G$ is similar to that in [16]. On the one hand, the behavior of $C_G$ ensures the decline of temperature to support the change of $T_G$ in figure 2. On the other hand, the change of sign in the heat capacity reflects the change in the stability of the thermodynamic system [16, 17], so the behavior of $C_G$ indicates the stability of black hole will get stronger before approaching the remnant mass. Therefore, $C_G$ is more comfortable than $C_A$ to describe a stable remnant.

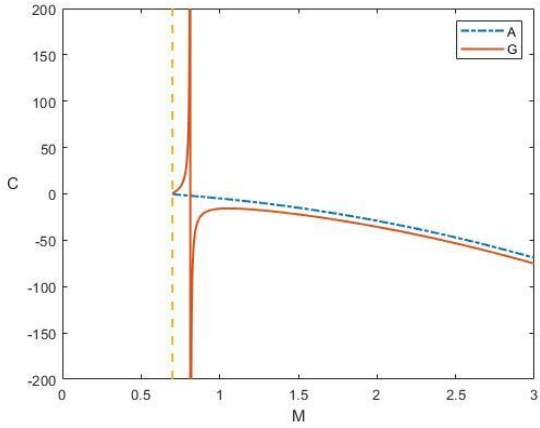

**Figure 3**. The heat capacity of black hole as given by Eq. (31) and Eq. (32) as the functions on $M$ for the original (A) and improved (G) methods when $\alpha > 0$.

In figure 4, $(dM/dt)_A$ increases to a finite value (It can be seen from Eq. (37)), which shows black hole arrives the highest evaporation rate at the critical case. On the contrary, $(dM/dt)_G$ first rises then fails rapidly, which shows black hole keeps a relatively low evaporation rate [18] before reaching the remnant mass. Referring to [15], the remnant is long-lived and owns a larger time to completely decay than the original black hole to decay to the remnant. For this reason, $(dM/dt)_G$ is accepted more easily than $(dM/dt)_A$ to construct a long evaporation process of the remnant.

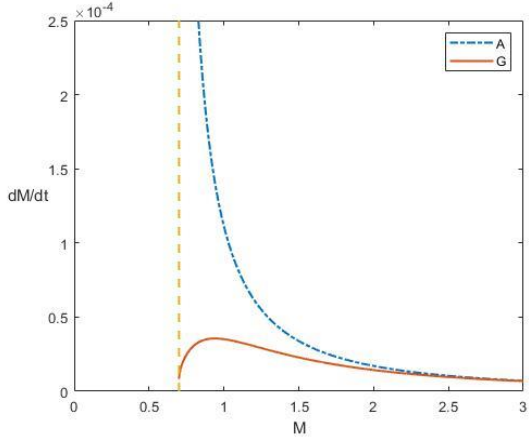

**Figure 4**. The evaporation rate of black hole as given by Eq. (37) and Eq. (38) as the functions on $M$ for the original (A) and improved (G) methods when $\alpha > 0$.

### 4.2 The evaporation with a negative parameter

In figure 5, it is shown how the sign of parameter $\alpha$ influences the generalized uncertainty principle. The positive $\alpha$ gives the minimum length for $\Delta x$, while the negative $\alpha$ lets $(\Delta x)_{min} = 0$ (The graph intersects the abscissa) to remove the minimum length. This means the generalized uncertainty principle is ineffective and physics becomes classical (uncorrected) again at Planck scale when the parameter is negative [37, 38]. Hence, in the following figures, the criterion for judging the rationality of figures should be: the negative parameter makes the thermodynamic quantities modified by the generalized uncertainty principle return to the original uncorrected case, and the changes of the thermodynamic quantities should follow the characteristics of uncorrected Hawking evaporation [1].

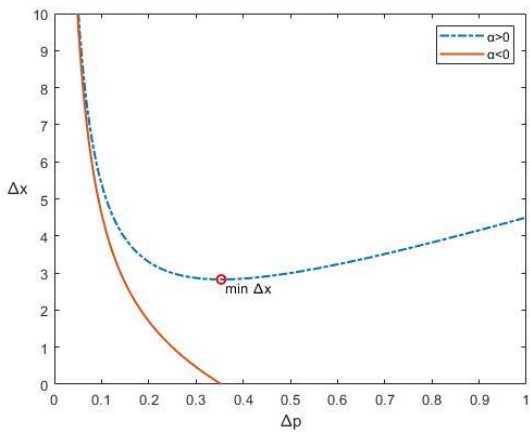

**Figure 5**. The $\Delta x$ as given by Eq. (3) as the functions on $\Delta p$ for $\alpha = 8$ and $\alpha =$

−8. The point circled in the figure is the minimum value for $\Delta x$.

In figure 6, both of $S_A$ and $S_G$ drops to 0 and the values of $S_A$ is generally bigger than that of $S_G$ in the whole process. When the mass disappears completely, the zero entropy is achieved for both $S_A$ and $S_G$, and the same uncorrected Hawking result is pointed out in [35]. There is no clear difference between the two entropies, but for other thermodynamic quantities that come from the differential of entropy, the difference will be amplified and their figures will be more distinct from each other.

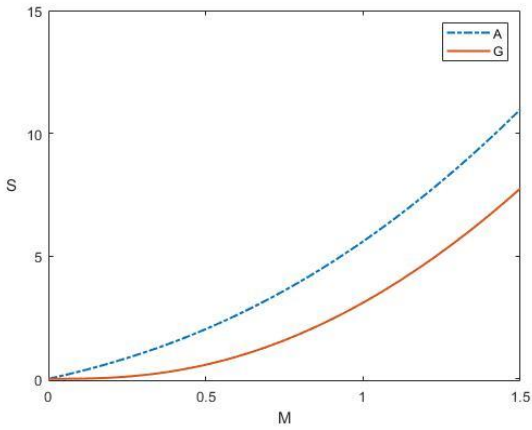

**Figure 6**. The entropy of black hole as given by Eqs. (24), (25) and (26) as the functions on $M$ for the original (A) and improved (G) methods when $\alpha < 0$.

In figure 7, when the evaporation ends, $T_A$ arrives to a finite nonzero value, while $T_G$ diverges leading to an infinite temperature that is usually derived from the uncorrected Hawking radiation ($T \propto 1/M$) [39]. According to the judgment criterion given by figure 5, the result of $T_G$ is consistent with uncorrected Hawking evaporation so it is reasonable. By contrast, although $T_A$ has a plausible eventual state [8], it is very strange to get a finite final temperature because the temperature in uncorrected Hawking evaporation is going to tend to infinity in the end [39].

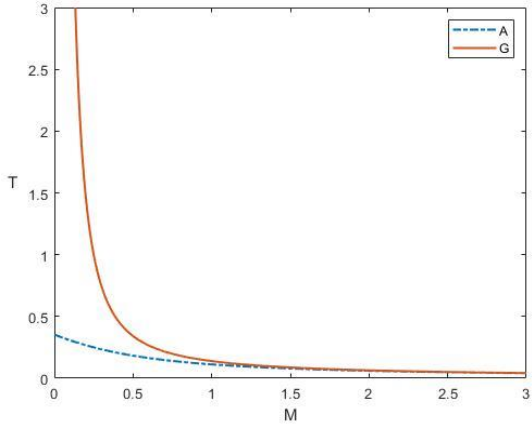

**Figure 7**. The temperature of black hole as given by Eq. (27) and Eq. (28) as the functions on $M$ for the original (A) and improved (G) methods when $\alpha < 0$.

In figure 8, as the mass gradually decays, both $C_A$ and $C_G$ go up more and more slowly. When the mass thoroughly disappears, $C_A$ reaches a negative value and $C_G$ grows to 0. The negative heat capacity reflects the thermodynamic system is unstable [40], and the zero heat capacity can prevent the black hole from evaporating [4]. For this reason, for the uncorrected Hawking evaporation, the final heat capacity should be 0. Thus, $C_A$ cannot represent the eventual evaporation phase. On the contrary, $C_G$ turns into 0 at the last moment, which is needed by a stable system.

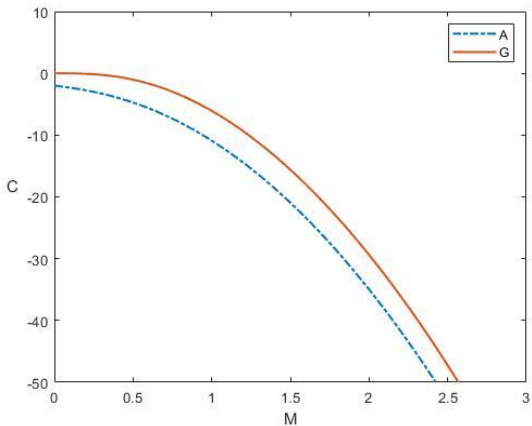

**Figure 8**. The heat capacity of black hole as given by Eq. (31) and Eq. (32) as the functions on $M$ for the original (A) and improved (G) methods when $\alpha < 0$.

In figure 9, $(dM/dt)_A$ goes up to a peak and down to 0, while $(dM/dt)_G$ gets larger and larger to infinity. According to the judgment criterion given by figure 5, the evaporation rate should follow the characteristics of uncorrected Hawking evaporation [1]: the black hole radiates faster and breaks down to its eventual disappearance. In other words, the slow evaporation will turn into a violent explosion leading to an infinite evaporation rate [39]. As we can see, $(dM/dt)_G$ is more comfortable than $(dM/dt)_A$ to describe the uncorrected Hawking evaporation because of its final divergence.

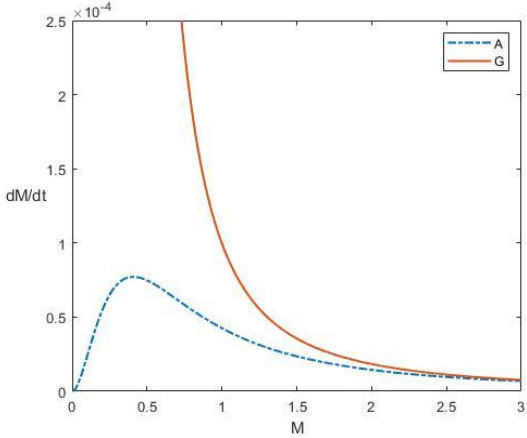

**Figure 9**. The evaporation rate of black hole as given by Eq. (37) and Eq. (38) as the functions on $M$ for the original (A) and improved (G) methods when $\alpha < 0$.

As $M \to 0$, Eq. (37) and Eq. (38) can be simplified as:

$$\left(\frac{dM}{dt}\right)_A \simeq -\frac{1}{60}\frac{M^2 c^2}{\pi \alpha^2 \hbar}, \tag{39}$$

$$\left(\frac{dM}{dt}\right)_G \simeq -\frac{1}{60 \times 16^5}\frac{\alpha^2 M_p^8 c^2}{\pi \hbar M^6}. \tag{40}$$

Here we add the minus sign to indicate the black hole is decaying. Solve the above functional equations to obtain:

$$M_A = M_0\left(\frac{\frac{60\pi\alpha^2\hbar}{c^2}}{\frac{60\pi\alpha^2\hbar}{c^2} + M_0 t}\right), \tag{41}$$

$$M_G = M_0\left(1 - \frac{7}{60 \times 16^5}\frac{\alpha^2 M_p^8 c^2}{\pi \hbar M_0^7} t\right)^{\frac{1}{7}}, \tag{42}$$

where $M_0$ is the initial mass at some point as $M \to 0$. By using Eq. (41) and Eq. (42), the functions of mass on time are painted in figure 10.

In figure 10, $M_A$ needs infinite time to complete the evaporation [8], while $M_G$ can arrive to 0 in a finite time. Considering the failure of the generalized uncertainty principle when $\alpha < 0$, the evaporation time should be finite, as the same as the result of uncorrected Hawking evaporation [41]. So $M_G$ is more appropriate to describe the final stage of black hole than $M_A$.

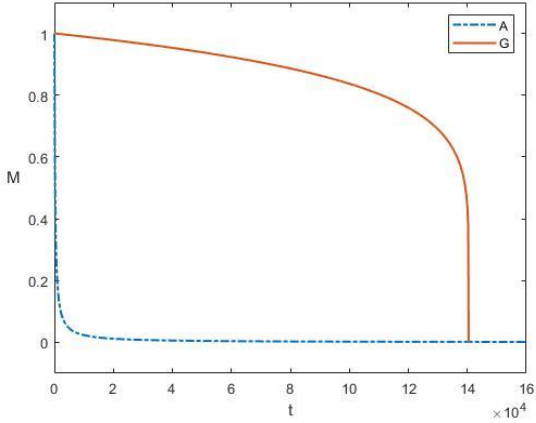

**Figure 10**. The mass of black hole as given by Eq. (41) and Eq. (42) as the functions on $t$ for the original (A) and improved (G) methods when $\alpha < 0$ and $M_0 = 1$.

From the two aspects for positive and negative parameters, we can see the figures of thermodynamic quantities from the improved method agree with the results described in many papers. On the contrary, there are some contradictions between the figures from the original method and these papers. In addition, both quantum gravity and non-commutative geometry, as same as the generalized uncertainty principle, reflect the existence of minimum length [42, 43], so the results from them can be the standard to measure the quality of methods for the generalized uncertainty principle. Because the figures of thermodynamic quantities respectively given by quantum gravity [12] and non-commutative geometry [13] are consistent with the figures from the improved method under a positive parameter, the improved method is proved to be reasonable again.

## 5. Conclusions

Based on the different treatments of the generalized uncertainty principle, we show two approximation methods to calculate thermodynamic quantities of black hole: one is the most common original method to change the generalized uncertainty principle into an approximate formula as Eq. (6), and the another is the improved method that we propose to replace Eq. (6) with Eq. (12). It is worth emphasizing that there is an essential difference between Eq. (6) and Eq. (12) even though they are so similar in form (see the discussions in Section. 2). Moreover, Eq. (21) is the key to distinguish between the applications of Eq. (6) and Eq. (12), so Eq. (21) also becomes the key to distinguish the improved method from the original method. Following the sign of parameter of the generalized uncertainty principle, we compare the two methods during the evaporation of black hole from positive and negative aspects: when the parameter is positive, the original method leads to an unstable remnant with high temperature and high evaporation rate, while the improved method gives a stable remnant with low temperature and low evaporation rate. Thus, the results of the improved method are more suitable to describe a long-lived remnant system that allows quantum gravity could resolve the information problem.

When the parameter is negative, the original method lets the black hole needs an infinite time to thoroughly decay with finite temperature and negative heat capacity. By contrast, the improved method makes the black hole complete the evaporation with infinite temperature and zero heat capacity in a finite time. Because the negative parameter can remove the minimum length to turn the thermodynamic quantities into the previous uncorrected case (It means the black hole has a finite lifetime and a divergent final temperature), the results of the improved method reflect the case of the negative parameter better.

From the above two aspects, we can see the improved method can explain the evaporation of black hole more reasonably than the original method. Meanwhile, the results from quantum gravity and non-commutative geometry also agree with the results from the improved method. However, there are still many articles to use the original method ignoring the huge impact it could have, so that we cannot judge whether these follow-up conclusions based on the original method make sense. Thus, we point out the problems brought by the original method, and provide a relatively reasonable improved method: to replace Eq. (6) with Eq. (12). Although we only apply the improved method for the black hole thermodynamic in this paper, the improved method can also be used for other thermodynamic systems by changing the treatments.